\documentclass[preprint]{imsart}
\RequirePackage[OT1]{fontenc}
\usepackage{amsthm,amsmath,natbib,bm,url}
\RequirePackage[colorlinks,citecolor=blue,urlcolor=blue]{hyperref}
\usepackage{booktabs}
\newcommand{\BF}{\mathrm{BF}}
\newcommand{\fbbf}{\mathrm{BF}^{F}_{(A+1):1}}
\newcommand{\fbbfB}{\mathrm{BF}^{F}_{(B+1):1}}
\newcommand{\fbbfAplusB}{\mathrm{BF}^{F}_{(A+B+1):1}}
\newcommand{\fbbfAB}{\mathrm{BF}^{F}_{(A+1)(B+1):1}}
\newcommand{\vs}{_{(A+1):1}}
\newcommand{\bicbf}{\mathrm{BF}^{\mathrm{BIC}}_{(A+1):1}}
\newcommand{\bicbfB}{\mathrm{BF}^{\mathrm{BIC}}_{(B+1):1}}
\newcommand{\bicbfAplusB}{\mathrm{BF}^{\mathrm{BIC}}_{(A+B+1):1}}
\newcommand{\bicbfAB}{\mathrm{BF}^{\mathrm{BIC}}_{(A+1)(B+1):1}}
\newcommand{\PR}{\mathrm{Pr}}

\startlocaldefs
\numberwithin{equation}{section}
\theoremstyle{plain}
\newtheorem{thm}{Theorem}[section]

\theoremstyle{remark}
\newtheorem{remark}{Remark}[section]

\endlocaldefs
\newcommand{\plim}{\mathop{\mathrm{plim}}}
\def\citeapos#1{\citeauthor{#1}'s (\citeyear{#1})}

\allowdisplaybreaks

\begin{document}

\begin{frontmatter}
\title{A Bayes factor with reasonable model selection consistency for ANOVA model}
\runtitle{A Bayes factor}

\begin{aug}
\author{\fnms{Yuzo} \snm{Maruyama}
\ead[label=e1]
{maruyama@csis.u-tokyo.ac.jp}}

\address{The University of Tokyo \\ \printead{e1}}

\runauthor{Y. Maruyama}

\end{aug}

\begin{abstract}
For the ANOVA model, we propose
a new g-prior based Bayes factor 
without integral representation, 
with reasonable model selection consistency 
for any asymptotic situations (either number of levels of the factor 
and/or number of replication in each level goes to infinity).
Exact analytic calculation of the marginal density 
under a special choice of the priors enables 
such a Bayes factor. 
\end{abstract}


\end{frontmatter}

\section{Introduction}
\label{sec:intro}
In this paper, we consider Bayesian model selection 
for ANOVA model.
We start with one-way ANOVA with two possible models. 
In one model,
all random variables have the same mean.
In the other model, there are some levels and
random variables in each level have different means.
Formally, the independent observations $y_{ij}$ 
($i=1,\dots,p, \ j=1,\dots,r_i, \ n=\sum_{i=1}^p r_i$) are assumed 
to arise from the linear model:
\begin{equation} \label{one-way-anova-model}
 y_{ij}=\mu+\alpha_i+\epsilon_{ij} , \quad \epsilon_{ij} \sim
 N(0,\sigma^2) 
\end{equation}
where $\mu$, $\alpha_i$ ($i=1,\dots,p$) and $\sigma^2$ are unknown. 
Clearly two models described above are written as follows:
\begin{equation} \label{model2}
\mathcal{M}_1: \ \alpha_i=0 \mbox{ for all }1\leq i \leq p, \
 \mbox{ vs }  \mathcal{M}_{A+1}: \ 
\alpha_i \neq 0 \mbox{ for some }i.
\end{equation}
Model selection, which refers to using the data in order to
decide on the plausibility of two or more competing models,
is a common problem in modern statistical science.
A natural Bayesian approach is by Bayes factor (ratio of 
marginal densities of two models), which is a function of
the posterior model probabilities (\cite{Kass-Raftery-1995}).

From a theoretical viewpoint, one of the most important topic on Bayesian model selection 
is consistency.
Consistency means that the true model will be chosen if enough data
are observed, assuming that one of the competing models is true.
It is well-known that BIC by \cite{Schwarz-1978} 
is consistent under fixed number of parameters.
In our situation, when $p$ is fixed and 
\begin{equation}
 \bar{r}=\frac{n}{p}=\frac{\sum_{i=1}^p r_i}{p}\to\infty,
\end{equation}
the BIC is consistent.

As a high-dimensional problem of statistical inference, 
the consistency in the case where $p \to \infty$ in one-way ANOVA setup, 
has been addressed by \cite{Stone-1979} and \cite{Berger-etal-2003}.
In the following, let ``CASE I'' and ``CASE II'' denote
the cases where 
\renewcommand{\theenumi}{\Roman{enumi}}
\begin{enumerate}
\item $n \to\infty$, $p\to\infty$ and $\bar{r}$ is bounded,
\item $n \to\infty$, $p\to\infty$ and $\bar{r}\to\infty$,
\end{enumerate}
respectively.
When $\sigma^2$ is known and CASE I is assumed, 
\cite{Stone-1979} showed that
BIC chooses the null model $\mathcal{M}_1$
with probability $1$ (that is, BIC is not consistent under $ \mathcal{M}_{A+1}$).
This is reasonable because BIC is originally derived by
the Laplace approximation under fixed number of parameters.
In the same situation as \cite{Stone-1979},
\cite{Berger-etal-2003} proposed a Bayesian criterion called GBIC,
which is derived by the Laplace approximation under CASE I.
Then they showed that GBIC has model selection consistency under CASE I.

Recently \cite{Moreno-etal-2010} showed that
under CASE II with $p=O(n^b)$ for $0<b<1$ or equivalently $ p=O(\{\bar{r}\}^{b/(1-b)})$,
BIC is consistent. By considering $b$ very close to $1$,
BIC seems to have consistency under CASE II
even if $ \bar{r}$ approaches infinity much slower than $p$.
As in Theorem \ref{main-thm-1}, however, 
we show that this is not so but the consistency of BIC under CASE II
depends on how quick $ \bar{r}$ goes to infinity compared to $p$.
As far as we know, this inconsistency
of BIC has not yet been reported in the literature. 

As an alternative to BIC, following \cite{Zellner-1986, Liang-etal-2008, Maruyama-George-2011},
we will propose a new $g$-prior based
Bayes factor with consistency under CASE II, which is given by
\begin{equation} \label{BFFB:1}
\fbbf
=\frac{\Gamma(p/2)\Gamma(\{n-p\}/2)}{\Gamma(1/2)\Gamma(\{n-1\}/2)}
\left( 1+\frac{W_H}{W_E}\right)^{(n-p-1)/2},
\end{equation}
where 
\begin{equation*}
\begin{split}
& W_H=\sum_{i}r_i(\bar{y}_{i\cdot}-\bar{y}_{\cdot \cdot})^2, \ 
W_E=\sum_{ij}(y_{ij}-\bar{y}_{i \cdot})^2, \\
&\bar{y}_{i \cdot}=
\frac{\sum_{j}y_{ij}}{r_i}, \ \bar{y}_{\cdot \cdot}=
\frac{\sum_{ij}y_{ij}}{\sum_i r_i}.
\end{split}
\end{equation*}
As shown in Theorem \ref{main-thm-1}, $\fbbf$ is consistent under CASE II
even if $ \bar{r}$ approaches infinity much slower than $p$.

Under CASE I where BIC is not consistent, 
the consistency of $\fbbf$ is shown to depend on a kind of distance
between $\mathcal{M}_{A+1}$ and $\mathcal{M}_1$, which is 
\begin{equation}\label{eq:cpr}
 c_A
=\frac{\sum_{ij}(E[y_{ij}]-E[\bar{y}_{\cdot\cdot}])^2}{n\sigma^2}.
\end{equation}
Naturally $c_A=0$ under $\mathcal{M}_1$ 
and $c_A>0 $ under $\mathcal{M}_{A+1}$.
In Theorem \ref{main-thm-1}, under CASE I with a positive small $ c_A$,
we show that $\fbbf$ chooses $\mathcal{M}_1$ with probability $1$.
Actually such an inconsistency result has been already
reported by \cite{Moreno-etal-2010}.
In Remark \ref{rem:inconsistency},
we demonstrate that the existence of inconsistency region 
for small $c_A$ and large $p$ is reasonable
from the prediction point of view.

The rest of the paper is organized as follows. In Section \ref{sec:BF},
we re-parameterize the ANOVA model
given by \eqref{one-way-anova-model} as a linear regression model and
give priors for the regression model. 
Then we derive marginal densities under 
$\mathcal{M}_1$ and $\mathcal{M}_{A+1}$ and eventually the Bayes factor
given by \eqref{BFFB:1}, as the ratio of the marginal densities.
In Section \ref{sec:consistency}, 
we show that the Bayes factor has a reasonable model
selection consistency for any asymptotic situation. 
In Section \ref{sec:two-way}, 
the results derived in Sections \ref{sec:BF} and \ref{sec:consistency}
are extended to two-way ANOVA model.
The Appendix presents some of the more technical proofs.

\section{A Bayes factor for one-way ANOVA}
\label{sec:BF}

\subsection{re-parameterized ANOVA}
\label{subsec:2.1}
Let $\bm{\alpha}=(\alpha_1,\dots,\alpha_p)'$, 
\begin{equation*}
\begin{split}
 \bm{y}&=(y_{11}, \dots , y_{1r_1}, y_{21}, \dots, y_{2r_2}, \dots, 
y_{p1},\dots,y_{pr_p})', \\
 \bm{\epsilon}&=(\epsilon_{11}, \dots , \epsilon_{1r_1}, \epsilon_{21}, \dots, \epsilon_{2r_2}, \dots, 
\epsilon_{p1},\dots,\epsilon_{pr_p})'. 
\end{split}
\end{equation*}
Further let an $n\times p$ matrix $\bm{X}_A=(x_{ij})$ satisfy
\begin{equation}
 x_{ij}=\begin{cases}
	 1 & \mbox{ if }\sum_{k=1}^{j-1}r_k+1 \leq i \leq \sum_{k=1}^{j}r_k, \\
	 0 & \mbox{otherwise}.
	\end{cases}
\end{equation}
Then the linear model given by \eqref{one-way-anova-model}
is written as
\begin{equation*}
 \bm{y}
=\mu \bm{1}_{n} + \bm{X}_A\bm{\alpha} +\bm{\epsilon} 
=\theta_1\bm{1}_{n} + \tilde{\bm{X}}_A\bm{\alpha} +\bm{\epsilon}
\end{equation*}
where $ \theta_1=\mu+\bm{r}'\bm{\alpha}/n$ with $\bm{r}=(r_1,\dots,r_p)'$
and 
\begin{equation*}
\tilde{\bm{X}}_A=
\bm{X}_A-\frac{\bm{1}_n \bm{r}'}{n}
\end{equation*}
which is the centered matrix of $\bm{X}_A$.
The matrix $\tilde{\bm{X}}_A$ is written as 
\begin{equation}
 (\overbrace{\bm{\nu}_1,\dots, \bm{\nu}_1}^{r_1},
\overbrace{\bm{\nu}_2,\dots, \bm{\nu}_2}^{r_2},\dots\dots
\overbrace{\bm{\nu}_p,\dots, \bm{\nu}_p}^{r_p})'
\end{equation}
where $ \bm{\nu}_i=\bm{e}_i -n^{-1}\bm{r}$ with
$\bm{e}_i=(0,\dots,0,1,0,\dots,0)'$.
The $n\times p$ matrix $\tilde{\bm{X}}_A$ 
is not full rank but $ \mbox{rank}\,\tilde{\bm{X}}_A=p-1$
since $ \{\bm{\nu}_1,\dots,\bm{\nu}_{p-1}\}$ is linearly independent and
$ \sum_{i=1}^{p}r_i\bm{\nu}_i=\bm{0}$.
For identifiability of $\bm{\alpha}$, 
the linear restriction 
\begin{equation}
 \bm{\nu}'_0\bm{\alpha}=0
\end{equation}
is assumed such that $\{\bm{\nu}_0, \bm{\nu}_1,\dots,\bm{\nu}_{p-1}\}$ is linearly independent.
Typically $\bm{1}_p$ or $\bm{r}$ are chosen as $\bm{\nu}_0$.
We will see that any linear restriction does not affect our result.

Let the singular value decomposition of
 $\tilde{\bm{X}}_A$ with rank $p-1$ be
\begin{equation*}
\tilde{\bm{X}}_A=
\bm{U}_A\bm{D}_A\bm{W}'_A, 
\end{equation*}
where $\bm{U}_A$ and $\bm{W}_A$ are $n \times (p-1)$ 
and $p \times (p-1)$ orthogonal matrices, respectively.
Then the one-way ANOVA is re-parameterized as the linear
regression model
\begin{equation}\label{re-parameterized-ANOVA}
\bm{y}
=\theta_1\bm{1}_{n} + \bm{U}_A\bm{\theta}_A +\bm{\epsilon}
\end{equation}
where
\begin{equation}
 \bm{\theta}_A=\bm{D}_A\bm{W}'_A\bm{\alpha}\in \mathcal{R}^{p-1}.
\end{equation}
Under the restriction $\bm{\nu}'_0\bm{\alpha}=0$, 
$ \bm{\theta}_A=\bm{0}_{p-1}$ is equivalent to $ \bm{\alpha}=\bm{0}_p$ because
the $p\times p$ matrix $(\bm{W}_A\bm{D}_A, \bm{\nu}_0)'$ is non-singular and
\begin{equation}
 \begin{pmatrix}
  \bm{D}_A\bm{W}'_A \\
\bm{\nu}'_0
 \end{pmatrix}\bm{\alpha}=
\begin{pmatrix}
 \bm{\theta}_A \\ 0
\end{pmatrix}.
\end{equation}
Hence, under \eqref{re-parameterized-ANOVA}, 
two models described in \eqref{model2} are written as
$ \mathcal{M}_1$: $ \bm{\theta}_A=\bm{0}_{p-1}$  and
$\mathcal{M}_{A+1}$: $\bm{\theta}_A \neq\bm{0}_{p-1} $.

\subsection{Priors and the Bayes factor}
\label{subsec:2.2}
In Bayesian model selection,
the specification of priors are needed on the models 
and parameters in each model.
For the former, let $ \PR(\mathcal{M}_1)=
       \PR(\mathcal{M}_{A+1})=1/2 $ as usual.
For the latter, at moment, 
we write joint prior densities as 
$p(\theta_1, \sigma^2)$ for $ \mathcal{M}_1$ and
$p(\theta_1, \bm{\theta}_A, \sigma^2)$ for $ \mathcal{M}_{A+1}$.
From the Bayes theorem,
$ \mathcal{M}_{A+1}$ is chosen when
$ \PR(\mathcal{M}_{A+1}|\bm{y}) > 1/2$ where
\begin{equation*}
\PR(\mathcal{M}_{A+1}|\bm{y})
=\frac{\BF\vs}{1+ \BF\vs}
\end{equation*}
and $\BF\vs$ is the Bayes factor given by
\begin{equation} \label{BF}
\BF\vs=m_{A+1}(\bm{y})/m_1(\bm{y}).
\end{equation}
In other words,
$ \mathcal{M}_{A+1}$ is chosen if and only if
$ \BF\vs  > 1$.
In \eqref{BF}, $m_\gamma(\bm{y})$ is 
the marginal density under $\mathcal{M}_\gamma$ for $\gamma=1, A+1$
as follows:
\begin{equation*}
\begin{split}
 m_1(\bm{y}) &= \iint p(\bm{y}|\theta_1,\sigma^2) 
p(\theta_1,\sigma^2)d\theta_1  d\sigma^2 \\
 m_{A+1}(\bm{y}) &= \iiint p(\bm{y}|\theta_1,\bm{\theta}_A,\sigma^2) 
p(\theta_1,\bm{\theta}_A,\sigma^2)d\theta_1 d\bm{\theta}_A d\sigma^2 ,
\end{split}
\end{equation*}
where $ p(\bm{y}|\theta_1,\sigma^2)$ and
$p(\bm{y}|\theta_1,\bm{\theta}_A,\sigma^2) $ are
sampling densities of $\bm{y}$ under $\mathcal{M}_1$ and
$\mathcal{M}_{A+1}$, respectively.

In this paper, we use the following 
joint prior density
\begin{equation} \label{prior-for-M1}
  p(\theta_1, \sigma^2) = 
 p(\theta_1) p(\sigma^2) = 1  \times  \sigma^{-2} 
\end{equation}
for $\mathcal{M}_1$ and
\begin{equation}\label{prior-for-MA+1}
 p(\theta_1,\bm{\theta}_A,\sigma^2) 
  =   p(\theta_1)  p(\sigma^2)  p(\bm{\theta}_A|\sigma^2) 
= \frac{1}{\sigma^2} 
 \int_0^\infty p(\bm{\theta}_A|g,\sigma^2)p(g)dg 
\end{equation}
for $\mathcal{M}_{A+1}$.
Note that $p(\theta_1) p(\sigma^2)=\sigma^{-2}$ in
both \eqref{prior-for-M1} and \eqref{prior-for-MA+1}
is  a popular non-informative prior. 
It is clear that
$\theta_1$ and $\sigma^2$  are location-scale parameters, and 
the improper prior for them is the right-Haar prior 
from the location-scale invariance group. 
These are reasons why 
the use of these improper priors for $\theta_1$ and $\sigma^2$ is 
formally justified. See \cite{Ber-Per-Var-1998, Berger-Bernardo-Sun-2009} 
for details.  

As $p(\bm{\theta}_A|g,\sigma^2)$, we  use so-called
\citeapos{Zellner-1986} $g$-prior 
\begin{equation}\label{eq:zellner-g}
p(\bm{\theta}_A|\sigma^2,g) = N_{p-1}(\bm{0}, g \sigma^2 (\bm{U}'_A\bm{U}_A)^{-1})
=N_{p-1}(\bm{0}, g \sigma^2 \bm{I}_{p-1}).
\end{equation}
As the prior of $g$, following \cite{Maruyama-George-2011},
we use Pearson Type VI distribution with the density
\begin{equation} \label{beta-prime}
 p(g)  = \frac{g^b(1+g)^{-a-b-2}}{B(a+1,b+1)} 
=\frac{g^{(n-p)/2-a-2}(1+g)^{-(n-p)/2}}{B(a+1,(n-p)/2-a-1)}
\end{equation}
where $-1<a<(n-p)/2-1$ and
\begin{equation}\label{beta-prime-1}
 b=(n-p)/2-a-2.
\end{equation}
\begin{remark}\label{rem:choice-a}
For the choice of $a$, my recommendation is $a=-1/2$.
We will describe it briefly.
The asymptotic behavior of $p(g) $ given by \eqref{beta-prime},
 for sufficiently large $g$, is proportional to $ g^{-a-2}$.
From the Tauberian Theorem, which is well-known for describing
the asymptotic behavior of the Laplace transform,
we have
\begin{equation}\label{tail-behavior-beta}
p(\bm{\theta}_A|\sigma^2)= \int_0^\infty p(\bm{\theta}_A|\sigma^2,g)p(g)dg 
\approx (\sigma^2)^{a+1} \|\bm{\theta}_A\|^{-(p+2a+1)},
\end{equation}
for sufficiently large 
$\bm{\theta}_A \in \mathcal{R}^{p-1}$, $a> -1$ and $b> -1$.
Hence 
the asymptotic tail behavior of $p(\bm{\theta}_A|\sigma^2)$ for $a=-1/2$
is multivariate Cauchy, $\|\bm{\theta}_A\|^{-(p-1)-1}$, 
which has been recommended by \cite{Zellner-Siow-1980} and others in
objective Bayes context.
\end{remark}

Before we proceed to give the Bayes factor with respect to priors described above,
we review statistical inference in ANOVA from frequentist point of view.
The key decomposition is given by
\begin{equation}\label{eq:decomp}
\begin{split}
 W_T &= \|\bm{y}-\bar{y}_{\cdot \cdot}\bm{1}_{n}\|^2=
\textstyle{\sum_{ij}}(y_{ij}-\bar{y}_{\cdot \cdot})^2 \\
&=\|\bm{U}_A\bm{U}'_A\{\bm{y}-\bar{y}_{\cdot \cdot}\bm{1}_{n}\}\|^2 
+\|(\bm{I}-\bm{U}_A\bm{U}'_A)\{\bm{y}-\bar{y}_{\cdot \cdot}\bm{1}_{n}\}\|^2 
\\
&= \textstyle{\sum_{i}}r_i(\bar{y}_{i\cdot}-\bar{y}_{\cdot \cdot})^2
+\textstyle{\sum_{ij}}(y_{ij}-\bar{y}_{i \cdot})^2 \\
&=
 W_H + W_E ,
\end{split}
\end{equation}
where $W_H$ and $W_E$ are independent and
\begin{equation*}
 \frac{W_H}{\sigma^2}\sim\chi^2_{p-1}\left[\|\bm{\theta}_A\|^2/\sigma^2\right], \ \frac{W_E}{\sigma^2}\sim\chi_{n-p}^2.
\end{equation*}
In \eqref{eq:decomp}, 
$W_T$, $W_H$ and $W_E$ are called
``total sum of squares'', ``between group sum of squares''
and ``within group sum of squares'', respectively.
The hypothesis $H_0:\bm{\alpha}=\bm{0}_p$ (or $\bm{\theta}_A=\bm{0}_{p-1}$)
is rejected if $ W_H/W_E$ is relatively larger.

In the theorem below, the Bayes factor under our priors
is an increasing function of $ W_H/W_E$, which is reasonable
from frequentist point of view, too.
\begin{thm}\label{thm:main-BF}
Under priors given by \eqref{prior-for-M1} under $\mathcal{M}_1$
and by \eqref{prior-for-MA+1} with \eqref{eq:zellner-g} and 
\eqref{beta-prime} under $\mathcal{M}_{A+1}$,
the Bayes factor is given by
 \begin{equation*}
 \frac{m_{A+1}(\bm{y}) }{m_1(\bm{y})}
=\frac{\Gamma(p/2+a+1/2)\Gamma((n-p)/2)}{\Gamma(a+1)\Gamma(\{n-1\}/2)}
\left(1+ \frac{W_H}{W_E}\right)^{(n-p-2)/2-a}.
\end{equation*}
\end{thm}
\begin{proof}
 See Appendix.
\end{proof}

By Remark \ref{rem:choice-a} and Theorem \ref{thm:main-BF},
the Bayes factor which we recommend is written as
\begin{equation} \label{BFFB}
\fbbf=\frac{m_{A+1}(\bm{y})}{m_1(\bm{y})}
=\frac{\Gamma(p/2)\Gamma(\{n-p\}/2)}{\Gamma(1/2)\Gamma(\{n-1\}/2)}
\left(1+ \frac{W_H}{W_E}\right)^{(n-p-1)/2},
\end{equation}
where the subscript F means ``Fully Bayes''. 

\section{Model selection consistency}
\label{sec:consistency}
In this section, we consider the model selection consistency as $n=\sum_i r_i\to\infty$.
Generally, the posterior consistency for model choice is defined as
\begin{equation}\label{def-plim}
 \plim_{n \to \infty}
\PR(\mathcal{M}_\gamma |y)=1 
\mbox{ or }
 \plim_{n \to \infty}
\BF_{\gamma':\gamma}=0
\end{equation}
when $ \mathcal{M}_\gamma $ is the true model and 
$ \mathcal{M}_{\gamma'} $ is not.
Here plim denotes convergence in probability and the probability
distribution in \eqref{def-plim}
is the sampling distribution under the true model $\mathcal{M}_\gamma$.
We will show that 
$\fbbf$ given by \eqref{BFFB}
has a reasonable model selection consistency.
As a natural competitor of $\fbbf$, the BIC based Bayes factor,
\begin{equation*}
\bicbf= \left(1+\frac{W_H}{W_E}\right)^{n/2}n^{-(p-1)/2},
\end{equation*}
is considered.

As remarked in Section \ref{sec:intro},
the consistency depends on $c_A$ given by \eqref{eq:cpr}
or equivalently $ \|\bm{\theta}_A\|^2/(n\sigma^2)$.
Let 
\begin{equation}
\tilde{c}_A=\lim_{n\to\infty}c_A
=\lim_{n\to\infty}\frac{\|\bm{\theta}_A\|^2}{n\sigma^2}.
\end{equation}
Then we have a following result.
\begin{thm} \label{main-thm-1}
\begin{enumerate}
 \item 
 Assume $\bar{r} \to \infty$ and $p$ is fixed.
\begin{enumerate}
 \item \label{main-thm-1-1}
$ \fbbf$ and $ \bicbf $ are consistent whichever the true model is.
\end{enumerate}
 \item  
Assume $p \to \infty$.
\begin{enumerate}
\item \label{main-thm-1-2}
$ \fbbf$ and $ \bicbf$ are consistent under $\mathcal{M}_1$.
\item 
Assume $\bar{r}$ is fixed under $\mathcal{M}_{A+1}$.
\begin{enumerate}
 \item \label{main-thm-1-3}
$ \fbbf$  is consistent (inconsistent)
if $\tilde{c}_A>(<) h(\bar{r})$ where
\begin{equation}\label{hr}
 h(r)= r^{1/(r-1)}-1.
\end{equation}
\item \label{main-thm-1-5}
$ \bicbf$  is inconsistent.
\end{enumerate}
\item 
Assume $\bar{r}\to\infty$ under $\mathcal{M}_{A+1}$.
\begin{enumerate}
 \item \label{main-thm-1-6}
$ \fbbf$ is consistent.
\item \label{main-thm-1-7}
$ \bicbf$ is consistent (inconsistent) if
$p\to\infty$ slower (faster) than 
$ (e\{1+\tilde{c}_A\}^{\bar{r}})/\bar{r}$. 
\end{enumerate}
\end{enumerate}
\end{enumerate}
\end{thm}
Note: \ $h(r)$ is a convex decreasing function in $r$ which satisfies
$h(2)=1$, $h(5) \doteq 0.5 $, $h(10) \doteq 0.29 $, and $h(\infty)=0$.
\begin{proof}
 See Appendix.
\end{proof}
Note: After \cite{Maruyama-2009}, the first version of this paper,
where the balanced ANOVA was treated only (that is, $r_i\equiv r$ is assumed), 
\cite{Wang-Sun-2012} extended consistency results of \cite{Maruyama-2009}
to unbalanced case while the proof itself essentially follows from \cite{Maruyama-2009}.

\begin{remark}\label{rem:inconsistency}
We give some remarks on inconsistency shown in Theorem \ref{main-thm-1}.
\begin{enumerate}
\item 
As shown in \ref{main-thm-1-5} of Theorem \ref{main-thm-1},
BIC always chooses $\mathcal{M}_1$ when $p \to \infty$ and $\bar{r}$ is fixed,
       even if $\tilde{c}_A$ is very large.
This is interpreted as unknown variance version of \citeapos{Stone-1979}.
\item 
As seen in \ref{main-thm-1-3} of Theorem \ref{main-thm-1}, 
$ \fbbf$ has an inconsistency region.
Actually existing such an inconsistency region has been also
reported by \cite{Moreno-etal-2010}.
They proposed intrinsic Bayes factor for normal regression model
and their upper-bound of
inconsistency region for one-way balanced ANOVA is given by
\begin{equation*}
\frac{r-1}{(r+1)^{(r-1)/r}-1}-1
\end{equation*}
which is slightly smaller than $h(r)$ given by \eqref{hr}.
\item
The existence of inconsistency region 
for small $c_A$ and large $p$ is quite reasonable from the
following reason.
Assume new independent observations $z_{ij}$ 
($i=1,\dots ,p$, $j=1,\dots,r_i$) from the same model as $y_{ij}$.
Then the difference of scaled mean squared prediction errors of
$\bar{y}_{i \cdot} $  and $\bar{y}_{\cdot \cdot} $ is given by
\begin{equation*}
\begin{split}
 \Delta[\bar{y}_{\cdot \cdot}; \bar{y}_{i \cdot}]
&=
\frac{E_{y,z}\left[
 \textstyle{\sum_{i,j}(z_{ij}-\bar{y}_{\cdot \cdot}
 )^2}\right]}{n\sigma^2}  
- \frac{E_{y,z}\left[
 \textstyle{\sum_{i,j}(z_{ij}-\bar{y}_{i \cdot}
 )^2}\right]}{n\sigma^2}  \\
&= c_A-\frac{p-1}{n} 
\end{split}
\end{equation*}
for any $p$ and $\bm{r}$.
First assume that $\mathcal{M}_1$ is true. 
We see that
\begin{equation*}
 \Delta[\bar{y}_{\cdot \cdot}; \bar{y}_{i \cdot}]= -(p-1)/n<0,
\end{equation*}
which is reasonable.
Then assume that $\mathcal{M}_{A+1}$ is true. 
When $\bar{r} \to \infty$ and $p$ is fixed,
\begin{equation*}
 \lim_{\bar{r} \to \infty}\Delta[\bar{y}_{\cdot \cdot}; \bar{y}_{i \cdot}]
=\tilde{c}_A>0,
\end{equation*}
which is reasonable.
On the other hand, if $p \to \infty$ and $\bar{r}$ is fixed,
\begin{equation*}
\lim_{p\to\infty}\Delta[\bar{y}_{\cdot \cdot}; \bar{y}_{i \cdot}]
= \tilde{c}_A-1/\bar{r}.
\end{equation*}
Hence when $ \tilde{c}_A< 1/\bar{r}$, 
$ \Delta[\bar{y}_{\cdot \cdot}; \bar{y}_{i \cdot}]$ is negative
even if $\mathcal{M}_{A+1}$ is true.
Therefore, from the prediction point of view, 
the existence of inconsistency region 
for small $c_A$ and large $p$ is reasonable.
\end{enumerate}
\end{remark}
Table \ref{tab:anova} shows frequency of choice of the true model
in some cases in numerical experiment where 
he balanced case $r_1=\dots=r_p$ is considered.
We see that it clearly guarantees the validity of Theorem
\ref{main-thm-1}. 

\begin{table} 
\caption{Frequency of choice of the true model}
\label{table1}
\begin{tabular}{ccccccccccccc} \toprule
 & $p\backslash r $ & 2 & 5 & 10 & 50 & 100 & 
& 2 & 5 & 10 & 50 & 100 \\ \midrule
& & \multicolumn{5}{c}{under $\mathcal{M}_1$} & \qquad & 
 \multicolumn{5}{c}{$c=0.1$ under  $\mathcal{M}_{A+1}$} \\ 
\cmidrule(lr){3-7} \cmidrule(lr){9-13}
FB & 2 &  0.77 & 0.89 & 0.94 & 0.98 & 0.98 & &  0.28 & 0.26 & 0.30 &
 0.80 & 0.97 \\
& 5 & 0.93 & 0.99 & 1.00 & 1.00 &1.00 &  & 0.19 & 0.09 & 0.15 & 0.78
 &1.00 \\
& 10 & 0.93 & 0.99 & 1.00 & 1.00 &1.00 & & 0.08 & 0.03 & 0.04 & 0.79
 &1.00 \\
& 50 &  1.00 & 1.00 & 1.00 & 1.00 &1.00 & & 0.00 & 0.00 & 0.00 & 0.91
 &1.00 \\
& 100 & 1.00 & 1.00 & 1.00 & 1.00 &1.00 & & 0.00 & 0.00 & 0.00 & 0.97
 &1.00 \\
\cmidrule(lr){3-7} \cmidrule(lr){9-13}
BIC & 2 & 0.53 & 0.79 & 0.91 & 0.97 & 0.97 & & 0.52 & 0.37 & 0.39 & 0.83
 & 0.98 \\
& 5 & 0.75 & 0.97 & 0.99 & 1.00 & 1.00 &  & 0.33 & 0.10 & 0.12 & 0.71 &
 0.99 \\
& 10 & 0.94 & 1.00  & 1.00 & 1.00 & 1.00 &  & 0.07 & 0.01 & 0.01 & 0.50
 & 0.99 \\
& 50 & 1.00 & 1.00 & 1.00 & 1.00 &1.00 &  & 0.00 & 0.00 & 0.00 & 0.00 &
 0.99 \\
& 100 & 1.00 & 1.00 & 1.00 & 1.00 &1.00 &  & 0.00 & 0.00 & 0.00 & 0.00 &
 0.99 \\  \midrule 

& & \multicolumn{5}{c}{$c=0.5$ under  $\mathcal{M}_{A+1}$} & \qquad & 
 \multicolumn{5}{c}{$c=1$ under  $\mathcal{M}_{A+1}$} \\ 
\cmidrule(lr){3-7} \cmidrule(lr){9-13}
FB & 2 & 0.48 & 0.66 & 0.86   &1.00& 1.00 &  & 0.68 & 0.88 &1.00 & 1.00
 & 1.00 \\
& 5 & 0.39 & 0.59 & 0.90   &1.00  &1.00 &  & 0.59 & 0.93 & 1.00  & 1.00
 & 1.00 \\
& 10 & 0.29 & 0.56 & 0.95   &1.00  &1.00 &  & 0.59 & 0.97 & 1.00  & 1.00
 & 1.00 \\
& 50 & 0.07 & 0.55 &1.00  &1.00  &1.00 &  & 0.52 & 1.00 & 1.00 &   1.00
 & 1.00  \\
& 100 & 0.03 & 0.55 &1.00  &1.00  &1.00 &  & 0.54 & 1.00 & 1.00 & 1.00
 & 1.00 \\
\cmidrule(lr){3-7} \cmidrule(lr){9-13}
BIC & 2 & 0.74 & 0.79 & 0.92 &1.00  &1.00 & & 0.87 & 0.95 & 1.00   &
 1.00   & 1.00 \\
& 5 & 0.57 & 0.61 & 0.88  &1.00  &1.00 &  & 0.79 & 0.94 & 1.00   & 1.00
 & 1.00  \\
& 10 & 0.26 & 0.29 & 0.78  & 1.00  &1.00 &  & 0.56 & 0.87 & 1.00  & 1.00
 & 1.00 \\
& 50 & 0.00 & 0.00 & 0.03 &  1.00  &1.00 &  & 0.00 & 0.03 & 1.00    &
 1.00   & 1.00 \\
& 100 & 0.00 & 0.00 & 0.00   &1.00  &1.00 &  & 0.00 & 0.00 & 1.00 &
 1.00   & 1.00 \\  \midrule 
& & \multicolumn{5}{c}{$c=2$ under  $\mathcal{M}_{A+1}$} & \qquad & 
 \multicolumn{5}{c}{$c=5$ under  $\mathcal{M}_{A+1}$} \\ 
\cmidrule(lr){3-7} \cmidrule(lr){9-13}
FB & 2 & 0.85 & 0.99 &   1.00  & 1.00   & 1.00  & & 0.98  & 1.00  & 1.00
 & 1.00  & 1.00  \\
& 5 & 0.87 & 1.00   & 1.00   & 1.00   & 1.00 &  & 1.00  & 1.00  & 1.00
 & 1.00  & 1.00 \\
& 10 & 0.89 & 1.00 &   1.00   & 1.00   & 1.00 &  & 1.00  & 1.00  & 1.00
 & 1.00  & 1.00 \\
& 50 & 1.00 & 1.00  & 1.00   & 1.00   & 1.00  &  & 1.00  & 1.00  & 1.00
 & 1.00  & 1.00 \\
& 100 & 1.00 & 1.00 &   1.00  & 1.00   & 1.00 & & 1.00  & 1.00  & 1.00
 & 1.00  & 1.00  \\
\cmidrule(lr){3-7} \cmidrule(lr){9-13}
BIC & 2 & 0.97 & 1.00  & 1.00  & 1.00  & 1.00 & & 1.00  & 1.00  & 1.00
 & 1.00  & 1.00 \\
& 5 & 0.97 & 1.00  & 1.00  & 1.00  & 1.00  &  & 1.00  & 1.00  & 1.00  &
 1.00  & 1.00 \\
& 10 & 0.88 & 1.00  & 1.00  & 1.00  & 1.00 &  & 1.00  & 1.00  & 1.00  &
 1.00  & 1.00 \\
& 50 & 0.02 & 1.00  & 1.00  & 1.00  & 1.00 &  & 0.89  & 1.00  & 1.00  &
 1.00  & 1.00 \\
& 100 & 0.00 & 0.92  & 1.00  & 1.00  & 1.00 & & 0.18  & 1.00  & 1.00  &
 1.00  & 1.00  \\  \bottomrule 
\end{tabular}
\label{tab:anova}
\end{table}

\section{two-way ANOVA}
\label{sec:two-way}
\subsection{re-parameterized two-way ANOVA}
\label{subsec:4.1}
In this section, we extend the results in Sections \ref{sec:BF} 
and \ref{sec:consistency} to two-way ANOVA model.
We have $ n$ independent normal random variables $y_{ijk}$
($i=1,\dots, p, \ j=1,\dots,q, \ k=1,\dots,r_{ij}, \ n=\sum_{i,j}r_{ij}$) 
where
\begin{equation} \label{two-way-anova-model}
 y_{ijk}= \mu + \alpha_i + \beta_j+ \gamma_{ij}+ 
\epsilon_{ijk}, \ \epsilon_{ijk} \sim  N(0,\sigma^2) .
\end{equation}
In the following, for simplicity as in \cite{Christensen-2011}, we assume
that $r_{ij}$ is given by
\begin{equation}\label{eq:rij}
 r_{ij}=\frac{n}{pq}\nu_i\xi_j
\end{equation}
where $\nu_i$ for $i=1,\dots,p$ and $ \xi_j$ for $j=1,\dots,q$ satisfy
\begin{equation*}
 \sum_{i=1}^p \nu_i=p, \ \sum_{j=1}^q\xi_j=q
\end{equation*}
and that $\{n/(pq)\}\nu_i\xi_j$ for any $(i,j)$ is a positive integer.
Therefore we allow some kinds of unbalanced design as well as balanced design,
$ r_{ij}\equiv n/(pq)$ for any $(i,j)$.

Let 
\begin{equation*}
\begin{split}
  \bm{y}&=(\{y_{111},\dots,y_{11r_{11}}\},\{y_{121},\dots,y_{12r_{12}}\}, \dots,\{y_{pq1},\dots,y_{pqr_{pq}}\})', \\
 \bm{\epsilon}&=(\{\epsilon_{111},\dots,\epsilon_{11r_{11}}\},\{\epsilon_{121},\dots,\epsilon_{12r_{12}}\}, \dots,\{\epsilon_{pq1},\dots,\epsilon_{pqr_{pq}}\})',
\end{split} 
\end{equation*}
$ \bm{\alpha}=(\alpha_1,\dots,\alpha_p)'$, $\bm{\beta}=(\beta_1,\dots,\beta_q)'$
and
\begin{equation*}
 \bm{\gamma}=(\{\gamma_{11},\dots,\gamma_{1q}\},
\{\gamma_{21},\dots,\gamma_{2q}\},\dots,\{\gamma_{p1},\dots,\gamma_{pq}\})'.
\end{equation*}
The $n\times (p+q+pq)$ design matrix is given by
\begin{equation*}
 \left(\bm{X}_A, \bm{X}_{B},\bm{X}_{AB}\right),
\end{equation*}
where the $n\times p$ matrix $\bm{X}_A=(x^A_{ti})$,
the $n\times q$ matrix $\bm{X}_B=(\bm{X}'_{B1},\dots,\bm{X}'_{Bp})'$ with 
an $(\sum_{j=1}^qr_{ij})\times q$ matrix $ \bm{X}_{Bi}=(x^{Bi}_{tj})$,
the $n\times pq$ matrix $\bm{X}_{AB}=(x^{AB}_{tm})$ 
satisfy
\begin{equation*}
 x_{ti}^A=\begin{cases}
	 1 & \mbox{ if }\sum_{m=1}^{i-1}\sum_{l=1}^q{r_{ml}}+1 \leq t \leq 
\sum_{m=1}^{i}\sum_{l=1}^q{r_{ml}}, \\
	 0 & \mbox{otherwise},
	\end{cases}
\end{equation*}
\begin{equation*}
 x_{tj}^{Bi}=\begin{cases}
	 1 & \mbox{ if }\sum_{m=1}^{j-1}r_{im}+1 \leq t \leq 
\sum_{m=1}^{j}r_{im}, \\
	 0 & \mbox{otherwise},
	\end{cases}
\end{equation*}
and
\begin{equation*}
 x_{tm}^{AB}=\begin{cases}
	 1 & \mbox{ if }\sum_{s< m}r_{i(s)j(s)}+1 \leq t \leq \sum_{s\leq m}r_{i(s)j(s)}, \\
	 0 & \mbox{otherwise},
	\end{cases}
\end{equation*}
where $i(s)$ and $j(s)$ are the unique integer solution of $s=(i-1)q+j$ where $i=1,\dots,p$ and $j=1,\dots,q$, respectively.
Then the linear model of \eqref{two-way-anova-model}
is written as
\begin{equation}\label{eq:two-way-anova-model-1}
 \bm{y}
=\mu \bm{1}_{n} + \bm{X}_A\bm{\alpha} +\bm{X}_B\bm{\beta}+\bm{X}_{AB}\bm{\gamma}+\bm{\epsilon} , \ \bm{\epsilon}\sim N_n(\bm{0}_n,\sigma^2\bm{I}_n).
\end{equation}
Note that $\{\mu, \bm{\alpha},\bm{\beta},\bm{\gamma}\}$ do not have
identifiability since
\begin{equation}
\begin{split}
& \bm{1}_n=\bm{X}_A\bm{1}_p=\bm{X}_{B}\bm{1}_q=\bm{X}_{AB}\bm{1}_{pq},  \\
& \bm{X}_A\bm{e}_i=\bm{X}_{AB}(\bm{0}_q,\dots,\bm{0}_q,\bm{1}_q,\bm{0}_q,\dots,\bm{0}_q), \mbox{ for } i=1,\dots,p, \\
&\bm{X}_B\bm{e}_j=\bm{X}_{AB}(\bm{e}'_j,\dots,\bm{e}'_j)', \mbox{ for }j=1,\dots,q .
\end{split} 
\end{equation}
So we will re-parameterize the model \eqref{eq:two-way-anova-model-1}
as a linear regression model with non-constrained parameters.

The expectation $E[\bm{y}]$ is re-written as
\begin{equation*}
\begin{split}
& \mu \bm{1}_{n} + \bm{X}_A\bm{\alpha} +\bm{X}_B\bm{\beta}+\bm{X}_{AB}\bm{\gamma} \\
&\quad =\theta_1(\mu,\bm{\alpha},\bm{\beta},\bm{\gamma})\bm{1}_n+\tilde{\bm{X}}_A\bm{\alpha} +\tilde{\bm{X}}_B\bm{\beta}+\tilde{\bm{X}}_{AB}\bm{\gamma}
\end{split}
\end{equation*}
where 
\begin{equation*}
\theta_1(\mu,\bm{\alpha},\bm{\beta},\bm{\gamma})=\mu+\frac{\bm{\nu}'\bm{\alpha}}{p}+\frac{\bm{\xi}'\bm{\beta}}{q}+
\frac{(r_{11},\dots,r_{pq})'\bm{\gamma}}{n}
\end{equation*}
and $\tilde{\bm{X}}_A, \ \tilde{\bm{X}}_B, \ \tilde{\bm{X}}_{AB}$
are the centered matrices.
As explained in \cite{Christensen-2011}, under the condition \eqref{eq:rij} on $r_{ij}$
including the balanced design, 
$\tilde{\bm{X}}_A$ and $ \tilde{\bm{X}}_B$ are orthogonal,
that is, $ \tilde{\bm{X}}'_A\tilde{\bm{X}}_B$ is the zero matrix.

Let the singular value decomposition of $\tilde{\bm{X}}_A$,
$\tilde{\bm{X}}_B$ and 
$(\bm{I}_n-\bm{U}_A\bm{U}'_A)(\bm{I}_n-\bm{U}_B\bm{U}'_B)\tilde{\bm{X}}_{AB} $ 
be 
\begin{equation}
\begin{split}
& \tilde{\bm{X}}_A =\bm{U}_A\bm{D}_A\bm{W}'_A ,\quad \tilde{\bm{X}}_B =\bm{U}_B\bm{D}_B\bm{W}'_B, \\
& (\bm{I}_n-\bm{U}_A\bm{U}'_A)(\bm{I}_n-\bm{U}_B\bm{U}'_B)
\tilde{\bm{X}}_{AB}  \\
&\qquad
=\bm{U}_{AB\backslash (A+B)}\bm{D}_{AB\backslash (A+B)}\bm{W}'_{AB\backslash (A+B)} ,
 \end{split}
\end{equation}
where the rank of $\tilde{\bm{X}}_A$, $\tilde{\bm{X}}_B$
and $(\bm{I}_n-\bm{U}_A\bm{U}'_A)(\bm{I}_n-\bm{U}_B\bm{U}'_B)\tilde{\bm{X}}_{AB}$ are $p-1$, $q-1$ and $(p-1)(q-1)$, respectively.
Then
\begin{equation*}
\begin{split}
 E[\bm{y}] &=\theta_1(\mu,\bm{\alpha},\bm{\beta},\bm{\gamma})\bm{1}_n
+\tilde{\bm{X}}_A\bm{\alpha}+\bm{U}_A\bm{U}'_A\bm{X}_{AB}\bm{\gamma} \\
& \qquad +\tilde{\bm{X}}_B\bm{\beta}+\bm{U}_B\bm{U}'_B\bm{X}_{AB}\bm{\gamma} 
+(\bm{I}_n-\bm{U}_A\bm{U}'_A-\bm{U}_B\bm{U}'_B)\tilde{\bm{X}}_{AB}\bm{\gamma} \\
&=\theta_1(\mu,\bm{\alpha},\bm{\beta},\bm{\gamma})\bm{1}_n
+\bm{U}_A\left\{\bm{D}_A\bm{W}'_A\bm{\alpha}+\bm{U}'_A\bm{X}_{AB}\bm{\gamma}\right\} \\
& \qquad +\bm{U}_B\left\{\bm{D}_B\bm{W}'_B\bm{\beta}+\bm{U}'_B\bm{X}_{AB}\bm{\gamma}\right\}
+(\bm{I}_n-\bm{U}_A\bm{U}'_A-\bm{U}_B\bm{U}'_B)\tilde{\bm{X}}_{AB}\bm{\gamma}\\
&=\theta_1(\mu,\bm{\alpha},\bm{\beta},\bm{\gamma})\bm{1}_n
+\bm{U}_A\bm{\theta}_A(\bm{\alpha},\bm{\gamma}) \\
&\qquad +\bm{U}_B\bm{\theta}_B(\bm{\beta},\bm{\gamma})
+\bm{U}_{AB\backslash (A+B)}\bm{\theta}_{AB\backslash (A+B)}(\bm{\gamma})
\end{split}
\end{equation*}
where
\begin{equation}
 \begin{split}
& \bm{\theta}_A(\bm{\alpha},\bm{\gamma})= \bm{D}_A\bm{W}'_A\bm{\alpha}+\bm{U}'_A\bm{X}_{AB}\bm{\gamma}\in \mathcal{R}^{p-1}, \\
& \bm{\theta}_B(\bm{\beta},\bm{\gamma})= \bm{D}_B\bm{W}'_B\bm{\beta}+\bm{U}'_B\bm{X}_{AB}\bm{\gamma}\in \mathcal{R}^{q-1}, \\
&\bm{\theta}_{AB\backslash (A+B)}(\bm{\gamma})=
\bm{D}_{AB\backslash (A+B)}\bm{W}'_{AB\backslash (A+B)}\bm{\gamma}\in
\mathcal{R}^{(p-1)(q-1)}.
 \end{split}
\end{equation}
Therefore the linear model with non-constrained parameters 
\begin{equation*}
 \{\theta_1,\bm{\theta}_A,\bm{\theta}_B, \bm{\theta}_{AB\backslash (A+B)}\}
\end{equation*}
is given by
\begin{equation}\label{eq:two-way-anova-model-2}
 \bm{y}=\theta_1\bm{1}_n+\bm{U}_A\bm{\theta}_A+\bm{U}_B\bm{\theta}_B
+\bm{U}_{AB\backslash (A+B)}\bm{\theta}_{AB\backslash (A+B)}+\bm{\epsilon}.
\end{equation}

In the two-way ANOVA model given by \eqref{eq:two-way-anova-model-1}, 
the following five submodels are important.
\begin{align*}
& \mathcal{M}_1: \ E[\bm{y}]=\mu\bm{1}_n, 
\quad (\bm{\alpha}=\bm{0}_p, \bm{\beta}=\bm{0}_q, \bm{\gamma}=\bm{0}_{pq}) \\
& \mathcal{M}_{A+1}: \ E[\bm{y}]=\mu\bm{1}_n+\bm{X}_A\bm{\alpha}, 
\quad (\bm{\beta}=\bm{0}_q, \bm{\gamma}=\bm{0}_{pq}) \\
& \mathcal{M}_{B+1}: \ E[\bm{y}]=\mu\bm{1}_n+\bm{X}_B\bm{\beta}, \quad
(\bm{\alpha}=\bm{0}_p, \bm{\gamma}=\bm{0}_{pq}) \\
& \mathcal{M}_{A+B+1}: \ E[\bm{y}]=\mu\bm{1}_n+\bm{X}_A\bm{\alpha}+\bm{X}_B\bm{\beta}, 
\quad (\bm{\gamma}=\bm{0}_{pq}) \\
& \mathcal{M}_{(A+1)(B+1)}: 
E[\bm{y}]=\mu\bm{1}_n+\bm{X}_A\bm{\alpha}+\bm{X}_B\bm{\beta}+\bm{X}_{AB}\bm{\gamma}.
\end{align*}

Under suitable constraints on $ \bm{\alpha}$, $\bm{\beta}$ and $\bm{\gamma}$,
for example,
\begin{equation}\label{eq:constraint_two_way}
 \bm{\nu}'\bm{\alpha}=0, \ \bm{\xi}'\bm{\beta}=0, \
\begin{pmatrix}
 r_{11},\dots,r_{pq} \\
\bm{U}'_A\bm{X}_{AB} \\
\bm{U}'_B\bm{X}_{AB}
\end{pmatrix}\bm{\gamma}=\bm{0}_{p+q-1},
\end{equation}
$ \{\mu,\bm{\alpha},\bm{\beta},\bm{\gamma}\}$ are identifiable, and further
between \eqref{eq:two-way-anova-model-1}
and \eqref{eq:two-way-anova-model-2}, there are the following equivalences:
\begin{align*}
& \mathcal{M}_1: \ 
 (\bm{\alpha}=\bm{0}_p, \bm{\beta}=\bm{0}_q, \bm{\gamma}=\bm{0}_{pq}) \\ &\qquad\qquad
\Leftrightarrow
(\bm{\theta}_A=\bm{0}_{p-1},\bm{\theta}_B=\bm{0}_{q-1},\bm{\theta}_{AB\backslash (A+B)}
=\bm{0}_{(p-1)(q-1)} ),\\
& \mathcal{M}_{A+1}: \ 
 (\bm{\beta}=\bm{0}_q, \bm{\gamma}=\bm{0}_{pq}) 
\Leftrightarrow
(\bm{\theta}_B=\bm{0}_{q-1},\bm{\theta}_{AB\backslash (A+B)}=\bm{0}_{(p-1)(q-1)} ), \\
& \mathcal{M}_{B+1}: \ 
 (\bm{\alpha}=\bm{0}_p, \bm{\gamma}=\bm{0}_{pq}) 
\Leftrightarrow
(\bm{\theta}_A=\bm{0}_{p-1},\bm{\theta}_{AB\backslash (A+B)}=\bm{0}_{(p-1)(q-1)} ), \\
& \mathcal{M}_{A+B+1}: \ 
  \bm{\gamma}=\bm{0}_{pq}
\Leftrightarrow
\bm{\theta}_{AB\backslash (A+B)}=\bm{0}_{(p-1)(q-1)} .
\end{align*}
Note that the constraints for $ \{\bm{\alpha},\bm{\beta},\bm{\gamma}\}$ as
\eqref{eq:constraint_two_way} do not affect our results.
\subsection{priors and the Bayes factor}
\label{subsec:4.2}

Following Section \ref{subsec:2.2}, we assume that
$\pi(\theta_1,\sigma^2)=1/\sigma^{2}$ for every model and
\begin{equation*}
 \bm{\theta}_A|\{g,\sigma^2\} \sim N_{p-1}(\bm{0}_{p-1},g\sigma^2\bm{I}_{p-1}) 
\end{equation*}
for $ \mathcal{M}_{A+1}, \mathcal{M}_{A+B+1}, \mathcal{M}_{(A+1)(B+1)}$,
\begin{equation*}
 \bm{\theta}_B|\{g,\sigma^2\} \sim N_{q-1}(\bm{0}_{q-1},g\sigma^2\bm{I}_{q-1}) 
\end{equation*}
for $ \mathcal{M}_{B+1}, \mathcal{M}_{A+B+1}, \mathcal{M}_{(A+1)(B+1)}$,
and
\begin{equation*}
 \bm{\theta}_{AB\backslash (A+B)}|\{g,\sigma^2\} \sim N_{(p-1)(q-1)}(\bm{0}_{(p-1)(q-1)},
g\sigma^2\bm{I}_{(p-1)(q-1)}) 
\end{equation*}
for $ \mathcal{M}_{(A+1)(B+1)}$.
Further we assume 
\begin{equation} \label{beta-prime-2}
 p(g)  = \frac{g^b(1+g)^{-a-b-2}}{B(a+1,b+1)} 
\end{equation}
where $a=-1/2$ and 
\begin{equation}\label{beta-prime-3}
 b(s)=(n-s-1)/2-a-2
\end{equation}
for 
\begin{align*}
 s= \begin{cases}
     p-1 & \mbox{ under }\mathcal{M}_{A+1}  \\
     q-1 & \mbox{ under }\mathcal{M}_{B+1}  \\
     p+q-2 & \mbox{ under }\mathcal{M}_{A+B+1} \\
     pq-1 & \mbox{ under }\mathcal{M}_{(A+1)(B+1)} .
    \end{cases}
\end{align*}
Before we proceed to give the Bayes factor with respect to priors described above,
we review the decomposition of squares for two-way ANOVA.
As similar in one-way ANOVA, the total sum of squares 
\begin{equation*}
 W_T=\sum_{ijk} (y_{ijk}- \bar{y}_{\cdot \cdot \cdot})^2
=\|\bm{y}-\bar{y}_{\cdot \cdot \cdot}\bm{1}_n \|^2
\end{equation*}
can be decomposed as 
\begin{equation}\label{eq:decomp-2way}
W_T= W_A+ W_B + W_{AB\backslash (A+B)} + W_E 
\end{equation}
where each sums of squares are given by
\begin{equation*}
\begin{split}
&W_A=\sum_{ijk} (\bar{y}_{i\cdot \cdot}- \bar{y}_{\cdot \cdot \cdot})^2
=\|\bm{U}_A\bm{U}'_A\{\bm{y}-\bar{y}_{\cdot \cdot \cdot}\bm{1}_n\} \|^2
=\|\bm{U}'_A\{\bm{y}-\bar{y}_{\cdot \cdot \cdot}\bm{1}_n\} \|^2, \\
&W_B=\sum_{ijk} (\bar{y}_{\cdot j \cdot}- \bar{y}_{\cdot \cdot \cdot})^2
=\|\bm{U}_B\bm{U}'_B\{\bm{y}-\bar{y}_{\cdot \cdot \cdot}\bm{1}_n\} \|^2
=\|\bm{U}'_B\{\bm{y}-\bar{y}_{\cdot \cdot \cdot}\bm{1}_n\} \|^2, \\
&W_{AB\backslash (A+B)}= \sum_{ijk} (\bar{y}_{ij\cdot} - \bar{y}_{i \cdot \cdot} 
- \bar{y}_{\cdot j \cdot} + \bar{y}_{\cdot \cdot \cdot})^2 \\
&=
\|\bm{U}_{AB\backslash (A+B)}\bm{U}'_{AB\backslash (A+B)}\{\bm{y}-\bar{y}_{\cdot \cdot \cdot}\bm{1}_n\} \|^2=
\|\bm{U}'_{AB\backslash (A+B)}\{\bm{y}-\bar{y}_{\cdot \cdot \cdot}\bm{1}_n\} \|^2, \\
&W_E = \sum_{ijk} (y_{ijk}- \bar{y}_{i j \cdot})^2  \\
&=\|(\bm{I}_n-\bm{U}_A\bm{U}'_A)(\bm{I}_n-\bm{U}_B\bm{U}'_B)
(\bm{I}_n-\bm{U}_{AB\backslash (A+B)}\bm{U}'_{AB\backslash (A+B)})
\{\bm{y}-\bar{y}_{\cdot \cdot \cdot}\bm{1}_n\} \|^2
\end{split}
\end{equation*}
and
\begin{equation*}
\begin{split}
& \bar{y}_{\cdot \cdot \cdot}=\frac{1}{n}\sum_{ijk}y_{ijk}, \quad
 \bar{y}_{i \cdot \cdot}=\frac{1}{\sum_j r_{ij}}\sum_{jk}y_{ijk}, \\
& \bar{y}_{ \cdot j \cdot}=\frac{1}{\sum_i r_{ij}}\sum_{ik}y_{ijk}, \quad 
 \bar{y}_{i j \cdot}=\frac{1}{r_{ij}}\sum_{k}y_{ijk} \ .
 \end{split}
\end{equation*}
In \eqref{eq:decomp-2way},
$ W_A, W_B,W_{AB\backslash (A+B)}$ and $W_E$ are independently distributed as
\begin{equation}
\begin{split}
& \frac{W_A}{\sigma^2}\sim \chi_{p-1}^2(\|\bm{\theta}_A\|^2/\sigma^2), \
 \frac{W_B}{\sigma^2}\sim \chi_{q-1}^2(\|\bm{\theta}_B\|^2/\sigma^2), \ 
 \frac{W_E}{\sigma^2}\sim \chi_{n-pq}^2, \\
& \frac{W_{AB\backslash (A+B)}}{\sigma^2}\sim \chi_{(p-1)(q-1)}^2(\|\bm{\theta}_{AB\backslash (A+B)}\|^2/\sigma^2).
\end{split} 
\end{equation}
Note that $ \|\bm{\theta}_A\|^2, \|\bm{\theta}_B\|^2$ and
$\|\bm{\theta}_{AB\backslash (A+B)}\|^2$ do not depend on parameterization
since these are given by
\begin{equation*}
\begin{split}
\|\bm{\theta}_A\|^2
&=\|\bm{U}'_A\{E[\bm{y}]-E[\bar{y}_{\cdot \cdot \cdot}]\bm{1}_n\} \|^2, \\
\|\bm{\theta}_B\|^2 
&=\|\bm{U}'_B\{E[\bm{y}]-E[\bar{y}_{\cdot \cdot \cdot}]\bm{1}_n\} \|^2, \\
\|\bm{\theta}_{AB\backslash (A+B)}\|^2 
&=\|\bm{U}'_{AB\backslash (A+B)}\{E[\bm{y}]-E[\bar{y}_{\cdot \cdot \cdot}]\bm{1}_n\} \|^2.
\end{split}
\end{equation*}
Based on the decomposition,
Bayes factors for two-way ANOVA are given as follows.
\begin{thm}\label{thm:two-way-main}
When 
$ \mathcal{M}_{\gamma} $  
for $\gamma=A+1, B+1, A+B+1, $ and $(A+1)(B+1)$
and $ \mathcal{M}_{1} $ are pairwisely compared,
the corresponding Bayes factors can be derived as follows:
\begin{align*}
& \fbbf
=\frac{\Gamma(p/2)\Gamma((n-p)/2)}{\Gamma(1/2)\Gamma(\{n-1\}/2)}
\left( 1+ \frac{W_A}{W_T-W_A}\right)^{(n-p)/2-1/2} \\
& \fbbfB=
\frac{\Gamma(q/2)\Gamma((n-q)/2)}{\Gamma(1/2)\Gamma(\{n-1\}/2)}
\left( 1+ \frac{W_B}{W_T-W_B}\right)^{(n-q)/2-1/2} \\
& \fbbfAplusB=
\frac{\Gamma(\frac{p+q-1}{2})\Gamma(\frac{n-p-q+1}{2})}{\Gamma(1/2)\Gamma(\{n-1\}/2)}
\left( 1+ \frac{W_A+W_B}{W_T-W_A-W_B}\right)^{(n-p-q)/2} \\
& \fbbfAB=
\frac{\Gamma(pq/2)\Gamma((n-pq)/2)}{\Gamma(1/2)\Gamma(\{n-1\}/2)}
\left( 1+\frac{W_T-W_E}{W_E}\right)^{(n-pq)/2-1/2} .
\end{align*}
\end{thm}

\subsection{Consistency}
\label{subsec:4.3}
In this subsection, we consider model selection consistency of Bayes factors
proposed in Theorem \ref{thm:two-way-main}.
We call $\{\mathrm{BF}^{F}\}$, the set of Bayes factors given in Theorem \ref{thm:two-way-main},
consistent under $\mathcal{M}_\gamma$ if and only if
\begin{equation*}
 \plim_{n\to\infty}\frac{\mathrm{BF}^{F}_{\gamma\prime:1}}{\mathrm{BF}^{F}_{\gamma:1}}
=0
\end{equation*}
for any $\gamma\prime\neq\gamma$ (Let $\mathrm{BF}^{F}_{1:1}=1$).

As the competitor, the corresponding Bayes factors based on BIC
are given as follows.
\begin{align*}
& \bicbf=n^{-(p-1)/2}
\left( 1+ \frac{W_A}{W_T-W_A}\right)^{n/2} \\
& \bicbfB=
n^{-(q-1)/2}\left( 1+ \frac{W_B}{W_T-W_B}\right)^{n/2} \\
& \bicbfAplusB=
n^{-(p+q-2)/2}
\left( 1+ \frac{W_A+W_B}{W_T-W_A-W_B}\right)^{n/2} \\
& \bicbfAB=n^{-(pq-1)/2}
\left( 1+\frac{W_T-W_E}{W_E}\right)^{n/2}.
\end{align*}

Let 
\begin{equation*}
 \tilde{c}_A=\lim_{n\to\infty}\frac{\|\bm{\theta}_A\|^2}{n\sigma^2}, \  
\tilde{c}_B=\lim_{n\to\infty}\frac{\|\bm{\theta}_B\|^2}{n\sigma^2}, \
 \tilde{c}_{AB\backslash (A+B)}=\lim_{n\to\infty}\frac{\|\bm{\theta}_{AB\backslash (A+B)}\|^2}{n\sigma^2}.
\end{equation*}
Then we have a following result on consistency.
\begin{thm}\label{thm:2way-main-consistency}
\begin{enumerate}
 \item  \label{2way-thm-1}
Assume $\bar{r} \to \infty$ and $p$ and $q$ are fixed.
\begin{enumerate}
 \item  \label{2way-thm-1-1}
Both $ \{\mathrm{BF}^{F}\}$ and $ \{\mathrm{BF}^{\mathrm{BIC}}\}$ are consistent whichever the true model is.
\end{enumerate}
\item   \label{2way-thm-2}
Assume $p \to \infty$ and $q \to \infty$. 
\begin{enumerate}
\item \label{2way-thm-2-1}
Both $ \{\mathrm{BF}^{F}\}$ and $ \{\mathrm{BF}^{\mathrm{BIC}}\}$ are consistent 
except under $\mathcal{M}_{(A+1)(B+1)}$.
\item \label{2way-thm-2-2}
Assume $\bar{r}$ is fixed.
\begin{enumerate}
 \item \label{2way-thm-2-2-1}
 $ \{\mathrm{BF}^{F}\}$ is consistent (inconsistent) under $\mathcal{M}_{(A+1)(B+1)}$ when
\begin{align}\label{two-way-condition}
\tilde{c}_{AB\backslash (A+B)}> (<) H(\bar{r},\tilde{c}_A+\tilde{c}_B),
\end{align}
where $H(r,c)$ with positive $c$ is the (unique) positive solution of
\begin{equation}
(x+1)^r/r-(x+1)-c=0.
\end{equation}
\item \label{2way-thm-2-2-2}
$ \{\mathrm{BF}^{\mathrm{BIC}}\}$ is inconsistent under $\mathcal{M}_{(A+1)(B+1)}$.
\end{enumerate}
\item  \label{2way-thm-2-3}
Assume $\bar{r}\to\infty$.
\begin{enumerate}
\item \label{2way-thm-2-3-1}
$ \{\mathrm{BF}^{F}\}$ is consistent under $\mathcal{M}_{(A+1)(B+1)}$.
\item \label{2way-thm-2-3-2}
$ \{\mathrm{BF}^{\mathrm{BIC}}\}$ is consistent (inconsistent)
under $\mathcal{M}_{(A+1)(B+1)}$ when
$pq\to\infty $ slower (faster) than $(e\{1+\tilde{c}_{AB\backslash (A+B)}\}^{\bar{r}})/\bar{r}$.
\end{enumerate}
\end{enumerate}
\end{enumerate}
\end{thm}
\begin{proof}
 See Appendix.
\end{proof}
The function $H(r,c)$ satisfies
\begin{itemize}
\item $ H(r,c)$ with fixed $r$ is increasing in $c$ 
and 
\begin{equation*}
 H(r,0)=h(r)
\end{equation*}
 where $h(r)$ is given by \eqref{hr}.
\item $ H(r,c)$ with fixed $c$ is decreasing in $r$ and 
\begin{equation*}
 \lim_{r\to\infty}H(r,c)=0.
\end{equation*}
\end{itemize}
\begin{remark}
We have results under fixed $q$ (and fixed $\bar{r}$ or $\bar{r}\to\infty$), too.
We omit the detail since they are somewhat complicated.
\end{remark}

\appendix
\section{Proof of Theorem \ref{thm:main-BF}}
First we derive the marginal density under $\mathcal{M}_1$.
Using the Pythagorean relation 
\begin{equation*}
 \| \bm{y} - \theta_1 \bm{1}_{n} \|^2
= n(\bar{y}_{\cdot \cdot}-\theta_1)^2 + W_T ,
\end{equation*}
where $\bar{y}_{\cdot \cdot}=n^{-1}\sum_{i,j}y_{ij} $ and
$W_T= \| \bm{y} - \bar{y}_{\cdot \cdot} \bm{1}_{n}  \|^2= 
\sum_{ij}(y_{ij}-\bar{y}_{\cdot \cdot})^2$,
we have
\begin{equation*}
\begin{split}
m_1(\bm{y}) &= 
 \int_{-\infty}^{\infty} \int_{0}^{\infty} 
\frac{1}{(2\pi)^{n/2}\sigma^{n+2}}
\exp\left(- \frac{\| \bm{y} - \theta_1 \bm{1}_{n} \|^2}
{2\sigma^2}\right)d\theta_1 d\sigma^2 \\
&= \frac{n^{1/2}}{(2\pi)^{n/2-1/2}}\int_{0}^{\infty} 
\frac{1}{\sigma^{n+1}}
\exp\left(- \frac{W_T}{2\sigma^2}\right) d\sigma^2 \\
&= \frac{n^{1/2}\Gamma(\{n-1\}/2)}{\pi^{(n-1)/2}} \{ W_T \}^{-(n-1)/2}.
\end{split} 
\end{equation*}

Then we derive the marginal density under $\mathcal{M}_{A+1}$.
Using the relationship
\begin{equation*}
\begin{split}
& \| \bm{y} - \theta_1 \bm{1}_{n} - \bm{U}_A\bm{\theta}_A\|^2
+g^{-1}\|\bm{\theta}_A\|^2 \\
& = n(\bar{y}_{\cdot \cdot}-\theta_1)^2 + 
\| \bm{y} - \bar{y}_{\cdot \cdot} \bm{1}_{n} -
 \bm{U}_A\bm{\theta}_A\|^2 +g^{-1}\|\bm{\theta}_A\|^2 \\
&= n(\bar{y}_{\cdot \cdot}-\theta_1)^2 + 
\frac{g+1}{g}\left\| \bm{\theta}_A - 
\frac{g\bm{U}'_A(\bm{y} - \bar{y}_{\cdot \cdot} \bm{1}_{n})}{g+1}\right\|^2 \\
&\qquad \quad - \frac{g}{g+1}\left\|\bm{U}'_A(\bm{y} - \bar{y}_{\cdot \cdot} \bm{1}_{n})\right\|^2 +\left\|\bm{y} - \bar{y}_{\cdot \cdot} \bm{1}_{n}\right\|^2 \\
&= n(\bar{y}_{\cdot \cdot}-\theta_1)^2 + 
\frac{g+1}{g}\left\| \bm{\theta}_A - 
\frac{g\bm{U}'_A(\bm{y} - \bar{y}_{\cdot \cdot} \bm{1}_{n})}{g+1}\right\|^2 
 + \frac{W_T+gW_E}{g+1},
\end{split}
\end{equation*}
where $W_E$ is given in \eqref{eq:decomp}, we have
the conditional marginal density of $\bm{y}$ given $g$ under $\mathcal{M}_{A+1}$,
\begin{equation}
\begin{split}
 m_{A+1}(\bm{y}|g) 
& = 
\int_{-\infty}^{\infty} \int_{\mathcal{R}^{p-1}} \int_{0}^{\infty} 
 p(\bm{y}|\theta_1,\bm{\theta}_A, \sigma^2)p(\bm{\theta}_A|\sigma^2,g)p(\sigma^2)
d \theta_1 d\bm{\theta}_A d\sigma^2  \\
&= \int_{-\infty}^\infty \int_{\mathcal{R}^{p-1}} \int_{0}^{\infty} 
\frac{1}{(2\pi\sigma^2)^{n/2}} \frac{1}{(2\pi g\sigma^2)^{(p-1)/2}}  \\
&  \qquad \times 
\exp\left( -\frac{\| \bm{y} - \theta_1 \bm{1}_{n} - \bm{U}_A\bm{\theta}_A\|^2}
{2\sigma^2}
-\frac{\|\bm{\theta}_A\|^2}{2g \sigma^2}
\right) 
p(\sigma^2)
d \theta_1 d\bm{\theta}_A d\sigma^2   \\
&= \int_{0}^{\infty} 
 \frac{n^{1/2}(1+g)^{-(p-1)/2}}{(2\pi\sigma^2)^{(n-1)/2}}
\exp\left( 
-\frac{W_T+gW_E}{2\sigma^2(g+1)}\right) 
\frac{1}{\sigma^2} d\sigma^2   \\
&= m_{1}(\bm{y}) 
\frac{(1+g)^{(n-p)/2}}{\left(g\{ W_E/W_T \}+1\right)^{(n-1)/2}}. 
\end{split}
\end{equation}
The fully marginal density 
\begin{equation*}
m_{A+1}(\bm{y})= \int_0^\infty m_{A+1}(\bm{y}|g) p(g)dg
\end{equation*}
under the prior of $g$ given by \eqref{beta-prime}, 
has a closed simple form as follows;
 \begin{equation*}
\begin{split}
 m_{A+1}(\bm{y}) 
&= m_1(\bm{y}) 
\int_0^\infty
\frac{(1+g)^{(n-p)/2}}{\left(g\{ W_E/W_T \}+1\right)^{(n-1)/2}}\frac{g^b(1+g)^{-a-b-2}}
{B(a+1,b+1)}dg \\
&= m_1(\bm{y}) \frac{1}{B(a+1,b+1)}
\int_0^\infty 
\frac{g^{b}}{\left(g\{ W_E/W_T \}+1\right)^{(n-1)/2}}dg \\
&=m_1(\bm{y})
\frac{\Gamma(p/2+a+1/2)\Gamma((n-p)/2)}{\Gamma(a+1)\Gamma(\{n-1\}/2)}
\left( \frac{W_T}{W_E}\right)^{(n-p-2)/2-a},
\end{split}
\end{equation*}
which completes the proof.

\section{Proof of Theorem \ref{main-thm-1}}
\label{app:B}
\begin{center}
 {\bfseries preparation: random part}
\end{center}
Recall that, for any $n$ and $p$, $W_H$ and $W_E$ are independent and
\begin{equation*}
 \frac{W_H}{\sigma^2}\sim\chi^2_{p-1}\left[\|\bm{\theta}_A\|^2/\sigma^2\right], \ \frac{W_E}{\sigma^2}\sim\chi_{n-p}^2.
\end{equation*}
In the following, $\stackrel{\mathrm{P}}{\to}_n $ denotes
convergence in probability as $n$ grows. 

When $p$ is fixed and $\bar{r}(=n/p)$ goes to $\infty$, we have
\begin{equation*}
\begin{split}
& \frac{W_E}{p\bar{r}\sigma^2}  \stackrel{\mathrm{P}}{\to}_{\bm{r}}1, \quad \frac{W_H}{\sigma^2}\sim\chi_p^2 \mbox{ under }\mathcal{M}_1, \\
& \frac{W_H}{p\bar{r}\sigma^2} 
\stackrel{\mathrm{P}}{\to}_{\bm{r}}\tilde{c}_A
\mbox{ under }\mathcal{M}_{A+1}.
\end{split}
\end{equation*}
Hence, for  any $\epsilon>0$ and any positive number $\bar{r}$,
there exists an $l(\epsilon)$ such that
\begin{equation}\label{r-infty-0}
 \PR\left(1/l(\epsilon)<\left\{1+W_H/W_E\right\}^{p\bar{r}}<l(\epsilon)\right)>1-\epsilon
\end{equation}
under $\mathcal{M}_1$. Under $\mathcal{M}_{A+1}$, we have
\begin{equation}\label{r-infty-1}
1+   \frac{W_H}{W_E} 
\stackrel{\mathrm{P}}{\to}_{\bm{r}}
1+\tilde{c}_A.
\end{equation}
When $p$ goes to infinity, we have
\begin{equation*}
 \begin{split}
& \frac{1}{\bar{r}-1}\frac{W_E}{p\sigma^2}  \stackrel{\mathrm{P}}{\to}_{p}1, \quad\frac{W_H}{p\sigma^2} \stackrel{\mathrm{P}}{\to}_{p}  1  \mbox{ under }\mathcal{M}_1, \\
& \frac{W_H}{p\sigma^2}  
\stackrel{\mathrm{P}}{\to}_{p}
1+\bar{r}\tilde{c}_A
 \mbox{ under }\mathcal{M}_{A+1}. 
 \end{split}
\end{equation*}
Hence we have
\begin{equation}\label{p-infty-1}
 \begin{split}
& 1+\frac{W_H}{W_E} \stackrel{\mathrm{P}}{\to}_{p} \frac{\bar{r}}{\bar{r}-1}\mbox{  under }\mathcal{M}_1, \\
& 1+  \frac{W_H}{W_E} 
\stackrel{\mathrm{P}}{\to}_{p}
\frac{\bar{r}\left\{1+\tilde{c}_A\right\}}{\bar{r}-1}
\mbox{ under }\mathcal{M}_{A+1}. 
 \end{split}
\end{equation}

\begin{center}
 {\bfseries preparation: non-random part}
\end{center}
For the asymptotic behavior of the gamma function,
 Stiring's formula,
\begin{equation} \label{Stiring}
\Gamma(ax+b) \approx \sqrt{2\pi}e^{-ax}(ax)^{ax+b-1/2}
\end{equation}
for sufficiently large $x$ is useful. 
Here $ f \approx g $ means $\lim f/g=1$.
Using \eqref{Stiring}, we get 
\begin{equation}\label{r-infty-2}
 \frac{\Gamma(p/2)\Gamma(\{ p\bar{r}-p\}/2)}{\Gamma(1/2)\Gamma(\{p\bar{r}-1\}/2)} 
 \approx  \frac{\Gamma(p/2)}{\Gamma(1/2)(p/2)^{(p-1)/2}}\frac{1}{\bar{r}^{(p-1)/2}},
\end{equation}
when $\bar{r} \to \infty$ and $p$ is fixed, and
\begin{equation} \label{p-infty-2}
\begin{split}
 \frac{\Gamma(p/2)\Gamma(\{ p\bar{r}-p\}/2)}{\Gamma(\{p\bar{r}-1\}/2)\Gamma(1/2)} 
 &\approx \sqrt{2}
 \frac{(\bar{r}-1)^{p(\bar{r}-1)/2-1/2}}{\bar{r}^{p\bar{r}/2-1}}\\
&= \frac{\sqrt{2}\bar{r}}{(\bar{r}-1)^{1/2}}\left\{
 \frac{\bar{r}-1}{\bar{r}^{\bar{r}/(\bar{r}-1)}}\right\}^{p(\bar{r}-1)/2}
\end{split}
\end{equation}
when $p \to \infty$.
\begin{center}
 {\bfseries main part}
\end{center}
First, we consider the consistency in the case where
 $r \to \infty$ and $p$ is fixed.
Using \eqref{r-infty-0}, \eqref{r-infty-1} and \eqref{r-infty-2}, we have
\begin{equation*}
 \fbbf \stackrel{\mathrm{P}}{\to}_{\bar{r}} 0 \mbox{ under }\mathcal{M}_1,
\quad
1/\fbbf
\stackrel{\mathrm{P}}{\to}_{\bar{r}} 0 \mbox{ under }\mathcal{M}_{A+1},
\end{equation*}
which means $\fbbf$ is consistent under both $\mathcal{M}_1$ and
$\mathcal{M}_{A+1}$.
Similarly we see that 
$\bicbf$ is consistent under both $\mathcal{M}_1$ and
$\mathcal{M}_{A+1}$ when $\bar{r}\to\infty$ and $p$ is fixed.

Then we consider the consistency in the case where 
$p \to \infty$.
Using \eqref{p-infty-1} under $\mathcal{M}_1$ and \eqref{p-infty-2}, 
we have
\begin{equation*}
\fbbf
\stackrel{\mathrm{P}}{\to}_p 0, \ \bicbf\stackrel{\mathrm{P}}{\to}_p 0
\end{equation*}
which means $\fbbf$ and $\bicbf$ 
are consistent under $\mathcal{M}_1$ when $p\to\infty$.

Using \eqref{p-infty-1} under $\mathcal{M}_{A+1}$ and \eqref{p-infty-2}, 
we have
\begin{equation*}
 \frac{1}{\fbbf}
\frac{\sqrt{2}\bar{r}}{(\bar{r}-1)^{1/2}}
\left\{\frac{1+\tilde{c}_A}{\bar{r}^{1/(\bar{r}-1)}}\right\}^{p(\bar{r}-1)/2}
\stackrel{\mathrm{P}}{\to}_p 1.
\end{equation*}
Hence when $\tilde{c}_A>h(\bar{r})=\bar{r}^{1/(\bar{r}-1)}-1$, we have
\begin{equation*}
 1/\fbbf\stackrel{\mathrm{P}}{\to}_p 0,
\end{equation*}
which means consistency of $\fbbf$ under $\mathcal{M}_{A+1}$.
Since $h(\bar{r})\to 0$ as $\bar{r}\to\infty$, $\fbbf$ has consistency
when both $p$ and $\bar{r}$ go to infinity.

On the other hand, when $\tilde{c}_A<h(\bar{r})$,
we have
\begin{equation*}
 \fbbf\stackrel{\mathrm{P}}{\to}_p 0,
\end{equation*}
which means inconsistency of $\fbbf$ under $\mathcal{M}_{A+1}$.

Using \eqref{p-infty-1} under $\mathcal{M}_{A+1}$ and \eqref{p-infty-2}, 
we have
\begin{equation*}
\bicbf
\left(p\bar{r}\left\{\frac{\bar{r}-1}{\bar{r}(1+\tilde{c}_A)}\right\}^{\bar{r}}\right)^{p/2}(p\bar{r})^{-1/2}
\stackrel{\mathrm{P}}{\to}_p 1.
\end{equation*}
We see that for any fixed $\bar{r}$, 
\begin{equation*}
\bicbf\stackrel{\mathrm{P}}{\to}_p 0
\end{equation*}
which means inconsistency of BIC under $\mathcal{M}_{A+1}$.
When $\bar{r}$ as well as $p$ goes to $\infty$, 
\begin{equation*}
\bicbf\left(\frac{p\bar{r}}{e(1+\tilde{c}_A)^{\bar{r}}}\right)^{p/2}(p\bar{r})^{-1/2}
\stackrel{\mathrm{P}}{\to}_p 1.
\end{equation*}
Hence if $p$ approaches infinity faster than 
$ (e\{1+\tilde{c}_A\}^{\bar{r}})/\bar{r}$, we have
\begin{equation*}
\bicbf\stackrel{\mathrm{P}}{\to}_{p,\bar{r}} 0,
\end{equation*}
which means inconsistency of BIC even if $\bar{r}\to\infty$.
On the other hand, 
if $p$ approaches infinity slower than 
$ (e\{1+\tilde{c}_A\}^{\bar{r}})/\bar{r}$, we have
\begin{equation*}
1/\bicbf \stackrel{\mathrm{P}}{\to}_{p,\bar{r}} 0.
\end{equation*}
which means consistency of BIC.

\section{Proof of Theorem \ref{thm:2way-main-consistency}}
By following the proof of Theorem \ref{main-thm-1} in Appendix \ref{app:B}, 
all of the proof are straightforward.
Here we give the sketch of the proof for part \ref{2way-thm-2-2-1} only.

When $p,q\to\infty$ and $\bar{r}$ is fixed,
\begin{equation}\label{eq:2way-pq}
\begin{split}
& \frac{W_A}{n\sigma^2} \stackrel{\mathrm{P}}{\to}_{p,q}\tilde{c}_A, \
\frac{W_B}{n\sigma^2} \stackrel{\mathrm{P}}{\to}_{p,q}\tilde{c}_B, \
\frac{W_E}{n\sigma^2} \stackrel{\mathrm{P}}{\to}_{p,q}1-1/\bar{r}, \\
&\frac{W_{AB\backslash (A+B)}}{n\sigma^2} \stackrel{\mathrm{P}}{\to}_{p,q}
\tilde{c}_{AB\backslash (A+B)}+1/\bar{r}.
\end{split}
\end{equation}
Using \eqref{eq:2way-pq}, Stiring's formula given in \eqref{Stiring} and
$\lim_{p\to\infty}p^{1/p}=1$,
we have
\begin{align*}
& \left\{\fbbf\right\}^{2/(pq)}
\stackrel{\mathrm{P}}{\to}_{p,q}
\left(\frac{1+\tilde{c}_A+\tilde{c}_B+\tilde{c}_{AB\backslash (A+B)}}{1+\tilde{c}_B+\tilde{c}_{AB\backslash (A+B)}}\right)^{\bar{r}}, \\
& \left\{\fbbfB\right\}^{2/(pq)}
\stackrel{\mathrm{P}}{\to}_{p,q}
\left(\frac{1+\tilde{c}_A+\tilde{c}_B+\tilde{c}_{AB\backslash (A+B)}}{1+\tilde{c}_A+\tilde{c}_{AB\backslash (A+B)}}\right)^{\bar{r}}, \\
& \left\{\fbbfAplusB\right\}^{2/(pq)}
\stackrel{\mathrm{P}}{\to}_{p,q}
\left(\frac{1+\tilde{c}_A+\tilde{c}_B+\tilde{c}_{AB\backslash (A+B)}}{1+\tilde{c}_{AB\backslash (A+B)}}\right)^{\bar{r}}, \\
& \left\{\fbbfAB\right\}^{2/(pq)}
\stackrel{\mathrm{P}}{\to}_{p,q}
\frac{(1+\tilde{c}_A+\tilde{c}_B+\tilde{c}_{AB\backslash (A+B)})^{\bar{r}-1}}{\bar{r}}.
\end{align*}
Therefore if
\begin{equation} \label{eq:2way-const-condition}
\frac{(1+ \tilde{c}_{AB\backslash (A+B)})^{\bar{r}}}{\bar{r}}>
1+\tilde{c}_A+\tilde{c}_B+\tilde{c}_{AB\backslash (A+B)},
\end{equation}
the ratio
\begin{equation} 
\left\{\frac{\mathrm{BF}^{F}_{\gamma:1}}{\fbbfAB}\right\}^{2/(pq)} ,
\end{equation}
for $\gamma=A+1,B+1,A+B+1$,
approaches a positive constant strictly less than $1$,
which guarantees the consistency of $ \fbbfAB$ 
under \eqref{eq:2way-const-condition}.

\end{document}